\documentclass[aps,onecolumn,prd,nofootinbib]{revtex4}
\usepackage{amsmath}
\usepackage{graphicx}
\usepackage{dcolumn}
\usepackage{bm}
\usepackage{amssymb}
\usepackage{latexsym}

\begin{document}

\title{Graviton mass generation in inspiralling double neutron stars}

\author{J. Wang\footnote{Email address: wangjing6@mail.sysu.edu.cn}}
\affiliation{School of Physics and Astronomy, Sun Yat-Sen University,
   Guangzhou, 510275, P. R. China}

\begin{abstract}
Although the Einsteins's general relativity predicted that the gravitational waves propagate at a speed of light and that the gravitons have no masses, many efforts, in keeping with some current cosmological observations, have been made to give graviton a mass and to realize a sound theory of massive gravity \cite{2012RvMP...84..671H, 2014LRR....17....7D}. However, it remains a mystery how gravitons generate a mass. According to Higgs mechanism, a spontaneous symmetry breaking is essential to a generation mechanism of property "mass" for associated particles. We point out that a spontaneous scalarized inspiring double neutron star (DNS) system can provide us a natural laboratory to investigate the generation mechanism of masses for gravitons. Because of the appearance of a gravitational scalar background field, with  small fluctuations, converged by iterative interplay of the mass dimensional external scalar fields, the binary system suffers from a spontaneous Lorentz symmetry breaking. The two scalarized NSs dip in a Higgs-like gravitational scalar potential, where the massless scalar background fluctuation field plays the role of Higgs field. Consequently, the gravitational scalar background field becomes massive. The radiated gravitons, propagating in a Yukawa-corrected potential, acquire a scalar-background-dependent mass term, in a massive-scalar-field-mediated way. We demonstrate that the mass of gravitons depends on intrinsic properties of the sources, which is not a certain value. The background-dependent masses for gravitons from scalarized orbital shrinking DNS is variable with the compactness of two components, as well as the separation of the binary. We get the effective masses for gravitons radiated from 8 detected DNS binaries with more precise mass measurements in our galaxy, whose values appear to be of the order of $10^{-23} ev/c^2$. It is found that more massive gravitons radiate from more closer DNS system, consisting with the higher-frequency gravitational waves from closer binaries.
\end{abstract}

\maketitle

\section{The scenario}

While it is remarked that the Einstein's general relativity has been so far the sound theory to describe the dynamics of neutron star (NS) binary systems, several observations indicated that the orbital decay in the double neutron star (DNS) system, PSR 1913+16, is mildly more rapid than that predicted by general relativistic quadrupole formula  \cite{1982ApJ...253..908T, 2010ApJ...722.1030W}. By making an analogy with the spontaneous magnetization of ferromagnets below the Curie temperature, a neutron star (NS), with a compactness of $\frac{Gm}{Rc^2}$\footnote{$m$ and $R$ are the mass and radius of NS, respectively. $G$ is the Newtonian gravitational constant.} above a certain critical value, will exhibit a nontrivial configuration, and a scalar field settles inside the NS \cite{1993PhRvL..70.2220D, 1996PhRvD..54.1474D}. That is to say a spontaneous scalarization occurs for the NS, which subsequently modifies its exterior space-time and produces a scalar asymptotic solution \cite{1998PhRvD..58l4003S}. As a result, an external scalar field $\varphi_1$ appears around the spontaneous scalarized NS, which triggers a scalarization of its companion and thus another external scalar field $\varphi_2$ around the induced scalarized NS\footnote{We assign the spontaneous scalarized NS as "NS 1" and the induced scalarized NS as "NS 2"}. The dynamical interplay between $\varphi_1$ and $\varphi_2$ is governed by a relation \cite{2014PhRvD..89d4024P}
\begin{equation}
^{(n+1)}\varphi_{i} = ^{(0)}\varphi_{i} + \frac{^{(n)}\varphi_{j}}{r},\label{fbm}
\end{equation}
where the indices $i, j = 1, 2$ denote two NSs, $^{(0)}\varphi_{i}$ is the initial external scalar field produced by the NS $i$, $^{(n)}\varphi^{i,j}$ represents the $n$th induced external scalar field around the NS $i$ or $j$, and $r$ is the distance from the centre of the binary. The feedback mechanism described by Eq. (\ref{fbm}) results in an iteratively induced scalarization between two NSs, which enhances the strength of the external scalar fields. Accordingly, a convergence of $^{(n)}\varphi_1$ and $^{(n)}\varphi_2$ occurs, leading to a gravitational scalar background field $\phi$. Therefore, the DNS is immersed in the gravitational scalar background field $\phi$, which has influence on the orbital dynamics of the binary system \cite{1992CQGra...9.2093D} and results in deviation from Einstein's general relativity \cite{1975ApJ...195L..51H, 1989ApJ...346..366W}.

In this letter, we consider a minimally coupled scalar-tensor theory and investigate the mechanism that the spin-2 gravitons generate a mass term due to the mediation of massive gravitational scalar field in scalarized DNS binaries. Instead of the usual method of effective field theory \cite{2003AnPhy.305...96A} that the graviton becomes heavy by eating Goldstone bosons because of the break of general coordinate invariance, we consider a potential, arising from the dynamical coupling between the gravitational scalar background field $\phi$ and the external scalar fields $\varphi_{i,j}$ of each NS, as well as the self-coupling of $\phi$. The iterative interplay between $\varphi_1$ and $\varphi_2$ causes small scalar fluctuations $\sigma$ ($\sigma \ll \phi$), which in turn imposes a conformal transformation on both the tensor metric and the gravitational scalar field $\phi$. By applying the conformal transformation to the tensor metric and scalar field that mediate the theory, the minimal coupling remains conformal invariant. The external background field $\varphi_{i,j}$ with mass dimension, under the conformal transformation, introduces a mass dimensional constant, which is responsible for a spontaneous symmetry breaking. As a consequent, the gravitational scalar field obtains a mass, in which the fluctuation field $\sigma$ plays the role of the Higgs field. The massive scalar field contributes to Yukawa type of corrections to the Newtonian potential of DNS binary, which has influences on the propagations of gravitational waves, endowing the gravitons with a mass.

\section{The Higgs mechanism in DNS}

The dynamics of a scalarized DNS binary is encoded not only by the gravitational tensor metric $g_{\mu\nu}$, but also by a gravitational scalar background field $\phi$, which naturally makes the scalar-tensor theory \cite{1992CQGra...9.2093D} of gravity to be the alternative theory to Einstein's general relativity describing the system. Neglecting the matter fields outside the NSs, we consider the scalar-tensor action that describes a DNS binary in vacuum,
\begin{equation}
S = \int d^4x \sqrt{-g} (\frac{M_{pl}^2}{2}\mathcal{R} - \frac{1}{2}g^{\mu\nu}\partial_{\mu}\phi\partial_{\nu}\phi - V(\phi)).\label{action}
\end{equation}
Here, $M_{pl} = \sqrt{1/8\pi G}$ is the reduced Planck constant. $\mathcal{R}$ and $g$ are the Ricci scalar and the determinant of the gravitational tensor metric $g_{\mu\nu}$, respevtively. $V(\phi)$ is the potential, consisting of the dynamical coupling of $\phi$ to $\varphi_{i,j}$ and a self-coupling term of $\phi$,
\begin{equation}
V(\phi) = \frac{\alpha}{2}\varphi_i\varphi_j\phi^2 + \frac{\lambda}{4}\phi^4.\label{potential}
\end{equation}
Here, $\alpha$ is a dimensionless coupling constant and characterizes the coupling strength between the scalarized NS and $\phi$, whose value depends on the compactness of NSs \cite{1992CQGra...9.2093D, 1993PhRvL..70.2220D, 1996PhRvD..54.1474D}. $\lambda$ is the self-coupling constant, which is roughly of the order of unity.
 
In addition, the iterative interplay and convergence between $^{(n)}\varphi_1$ and $^{(n)}\varphi_2$ perturb $\phi$ and cause small scalar background fluctuations $\sigma$ ($\sigma \ll \phi$), which also interacts with the dynamical fields, i.e. the gravitational tensor metric and the gravitational background scalar field, following the couplings \cite{1993PhRvL..70.2220D, 1998PhRvD..58l4003S, 2014PhRvD..89d4024P},
\begin{eqnarray}
g_{\mu\nu}^* &=& e^{-2\lambda\sigma}g_{\mu\nu},~~~\sqrt{-g^*} = e^{4\lambda\sigma}\sqrt{-g},\label{nm}\\
\phi^* &=& e^{-\lambda\sigma}\phi,\label{nscalar}
\end{eqnarray}
where $g_{\mu\nu}^*$, $g^*$, and $\phi^*$ are transformed gravitational tensor metric and its determinant, and transformed gravitational background scalar field.
By expanding the transformed metric $g_{\mu\nu}^*$ about a Minkowski background in terms of Eq. (\ref{nm}), we express them as
\begin{equation}
g_{\mu\nu}^* = \eta_{\mu\nu} + h_{\mu\nu}^*,~~~h_{\mu\nu}^* = h_{\mu\nu} + 2\eta_{\mu\nu}\lambda\sigma,\label{enm}
\end{equation}
where $|h_{\mu\nu}|, |h_{\mu\nu}^*| \ll 1$. The Eq. (\ref{nm}) remains unchanged.
According to Eq. (\ref{nm}) and Eq. (\ref{nscalar}), we find that the kinetic term in action (\ref{action}) is transformed into a canonical kinetic term,
\begin{eqnarray}
&&-\frac{1}{2}\sqrt{-g}g_{\mu\nu}\partial_{\mu}\phi\partial_{\nu}\phi = -\frac{1}{2}\sqrt{-g^*}g_{\mu\nu}^*\mathcal{D}_{\mu}\phi^*\mathcal{D}_{\nu}\phi^*,\\
&&\mathcal{D}_{\mu} \equiv \partial_{\mu} + \lambda\partial_{\mu}\sigma,
\end{eqnarray}
which is scale-invariant. The transformed action then reads
\begin{equation}
S^* =  \int d^4x \sqrt{-g^*} (\frac{M_{pl}^2}{2}\mathcal{R}^* - \frac{1}{2}g^{*\mu\nu}\mathcal{D}_{\mu}\phi^*\mathcal{D}_{\nu}\phi^* - V(\phi^*)).\label{naction}
\end{equation}

The solutions of external scalar fields $\varphi_{i,j}$ for each component have mass dimensions \cite{1998PhRvD..58l4003S}. In the process of conformal transformation, the solution with mass dimension involves a dimensional constant $\mu$, which appears in the transformed scalar potential $V(\phi^*)$,
\begin{equation}
V(\phi^*) = \frac{\alpha}{2}\mu^2\phi^{*2} + \frac{\lambda}{4}\phi^{*4}.\label{tpotential}
\end{equation}
The Planck scale constant $\mu = \sqrt{1/8\pi G_{\it eff}}$ appears to be related to the scalar charges of two scalarized NSs by the effective gravitational constant $G_{\it eff}$ \cite{1992CQGra...9.2093D}. It is the appearance of the mass-dimensional constant $\mu$ that contributes to a process of spontaneous breaking of symmetry, which allows us here to apply the similar recipe to the Higgs mechanism in the standard model.

Actual NSs observed in DNS binaries, with important deviations from general relativity in strong-field regime, would develop strong scalar charges in the absence of an external scalar field for enough negative values of $\alpha$, i.e. $\alpha < 0$ \cite{1993PhRvL..70.2220D, 1996PhRvD..54.1474D, 2014PhRvD..89d4024P}. The self-coupling constant $\lambda$ is of the order of unity, i.e. $\lambda > 0$.
By considering that the interplay between $\varphi_{i,j}$ is long-range force, the behavior of the gravitational background scalar field $\phi^*$ near spatial infinity endows it with a vacuum expectation value (VEV) $v_{\phi^*}$,
\begin{equation}
v_{\phi^*}^2 = -\frac{\alpha\mu^2}{2\lambda},\label{VEV}
\end{equation}
which is obtained from the condition $\frac{dV(\phi^*)}{d\phi^*}|_{\phi^*_{\it min}} = 0$. 
Therefore, we write the gravitational scalar background field $\phi^*$ as the combination of its VEV $v_{\phi^*}$ and the fluctuating field $\tilde{\phi^*}$ of the spatial infinity approximate value of $\phi^*$, according to
\begin{equation}
\phi^* = v_{\phi^*} + \tilde{\phi^*}.\label{escalar}
\end{equation}
Substituting the VEV (\ref{VEV}) and Eq. (\ref{escalar}) into the Lagrangian of $\phi^*$ extracted from Eq. (\ref{naction}), we can get the mass of $\phi^*$,
\begin{equation}
m_s^2 = -\alpha\mu^2.\label{smass}
\end{equation}

\section{The scalar background-dependent mass of gravitons}

Variation of the transformed action (\ref{naction}), with respect to the transformed metric (\ref{nm}) and the transformed scalar background field (\ref{nscalar}), yields the following equations of motion (e.o.m.) in vacuum,
\begin{eqnarray}
&&(\mathcal{R}^*_{\mu\nu}-\frac{1}{2}g_{\mu\nu}^*\mathcal{R}^*)M_{pl}^2 = \partial_{\mu}\phi^*\partial_{\nu}\phi^*-\frac{1}{2}g_{\mu\nu}^*(\partial_{\alpha}\phi^*)^2+\frac{\alpha}{2}\mu^2g_{\mu\nu}^*\phi^{*2}+\frac{\lambda}{4}g_{\mu\nu}^*\phi^{*4},\label{meom}\\
&&\Box_{g^*} \phi^* = \alpha\mu^2\phi^* + \lambda\phi^{*3},\label{seom}
\end{eqnarray}
where $\Box_{g^*}$ is the curved space d'Alembertian that is defined by $\Box_{g^*} = \sqrt{-g^*}\partial_{\nu}(\sqrt{-g^*}g_{\mu\nu}^*\partial_{\mu})$.

Let us expand the transformed scalar potential $V(\phi^*)$ of Eq. (\ref{tpotential}) in a Taylor series about the VEV of $\phi^*$ (\ref{VEV}),
\begin{equation}
V(\phi^*) = V(v_{\phi^*}) + V(v_{\phi^*})'\tilde{\phi^*} + \frac{1}{2}V(v_{\phi^*})''\tilde{\phi^*} + \cdots.\label{epotential}
\end{equation}
Considering the weak field scalar perturbation of $\phi^*$ in Eq. (\ref{escalar}), we expand the field equations in weak field limit,
\begin{eqnarray}
&&(\mathcal{R}^*_{\mu\nu}-\frac{1}{2}g_{\mu\nu}^*\mathcal{R}^*)M_{pl}^2 = \partial_{\mu}\phi^*\partial_{\nu}\phi^*-\frac{1}{2}g_{\mu\nu}^*(\partial_{\alpha}\phi^*)^2+\frac{1}{2}m_s^2(\phi^*-\phi_0)^2g_{\mu\nu}^*,\label{emeom}\\
&&(\Box_{g^*} - m_s^2) \phi^* = m_s^2v_{\phi^*}.\label{eseom}
\end{eqnarray}
Here, we use the expansion Eq. (\ref{epotential}) and just consider the leading-order terms. The expanded field equations that are consistent at all orders in $(\frac{v}{c})^n$ is required. We solve the e.o.m. of massive scalar field Eq. (\ref{eseom}) in a static spherically symmetric configuration, which yields an exterior solution of $\phi^*$ far from the DNS system,
\begin{equation}
\phi^* =v_{\phi^*} + v_{\phi^*}G_{\it eff}\frac{M_r}{r}e^{-m_sr},\label{ssolution}
\end{equation}
where $M_r = \frac{m_1m_2}{m_1+m_2}$ is the reduced mass of DNS, and $m_1$ and $m_2$ are the masses of two NSs in the binary. Therefore, the mass $m_s$ is the ordinary mass parameter in the Klein-Gordon equation (\ref{eseom}) of $\phi^*$, arising from the Higgs-like potential (\ref{tpotential}). In the meanwhile, the mass of $\phi^*$ plays precisely a role in the Yukawa-like correction $\sim e^{-m_sr}$ to the standard Newtonian form of gravitational potential $\sim \frac{GM_r}{r}$ in the scalarized DNS binary system.

We expand the left-hand side of Eq. (\ref{emeom}) in weak field limit, by using the weak field perturbations of tensor metric Eq. (\ref{enm}) and the small perturbative coupling $\theta^{\mu\nu} = h^{\mu\nu*} - \frac{1}{2}h^*\eta^{\mu\nu} - \frac{\tilde{\phi^*}}{v_{\phi^*}}\eta^{\mu\nu}$, with $h^* = \eta^{\mu\nu}h_{\mu\nu}^*$. Imposing the harmonic gauge $\partial^{\nu}(h_{\mu\nu}^*-\frac{1}{2}\eta_{\mu\nu}h^*) = 0$ and $\partial^{\nu}{\theta_{\mu\nu}} = 0$, and neglecting the higher-order terms, we rewrite the e.o.m. of gravitons as
\begin{eqnarray}
&&-\frac{M_{pl}^2}{2}\Box_{\eta}\bar{h}_{\mu\nu}^* - \frac{M_{pl}^2}{2}\Box_{\eta}\theta_{\mu\nu} - M_{pl}^2\eta_{\mu\nu}\Box_{\eta}(\frac{\tilde{\phi^*}}{v_{\phi^*}})\nonumber\\ &&= \partial_{\mu}\phi^*\partial_{\nu}\phi^*-\frac{1}{2}\eta_{\mu\nu}(\partial_{\alpha}\phi^*)^2+\frac{1}{2}m_s^2(\phi^*-v_{\phi^*})^2\eta_{\mu\nu},\label{geom}
\end{eqnarray}
where $\bar{h}_{\mu\nu}^* = h_{\mu\nu}^*-\frac{1}{2}\eta_{\mu\nu}h^*$ and the flat-space d'Alembertian $\Box_{\eta} = \eta^{\mu\nu}\partial_{\mu}\partial_{\nu}$.
Let us study the scalar-mediated propagations of gravitons outside the scalarized DNS binary. Substituting the solution of $\phi^*$ (\ref{ssolution}) into the e.o.m. of gravitons (\ref{geom}), we then write the wave solution of gravitons (\ref{geom}) as \cite{Maggiore:2008}
\begin{equation}
\bar{h}_{\mu\nu}^* = (\int d\omega\int\frac{d^3k}{(2\pi)^3}Ae^{i(\vec{k}\cdot\vec{r}-\omega t)})\cdot\sum\limits_{n=-\infty}^{\infty}\phi^{*n}(\vec{r})e^{in\sigma}.\label{tsolution}
\end{equation}
Here, $A$ denotes the amplitude of tensor gravitational waves radiated from the orbital decaying DNS. $\sum\limits_{n=-\infty}^{\infty}\phi^{*n}(\vec{r})e^{in\sigma}$ is the Fourier expansion of the gravitational scalar field $\phi^*$, with the gravitational background scalar fluctuation fields $\sigma$, which is converged by the $n$th induced scalar background $^{(n)}\varphi_{i,j}$ of two NSs. $\phi^{*n}$ is in the form of exterior solutions (\ref{ssolution}). 

The Klein-Gordon equation of gravitons (\ref{tsolution}) therefore reads
\begin{equation}
[\Box - \frac{m_s^2v_{\phi^*}G_{\it eff}}{\tilde{R}}]\bar{h}_{\mu\nu}^* = 0,\label{hKG}
\end{equation}
where $\tilde{R}$ is the semi-major axis of the elliptical DNS binary system. By defining
\begin{equation}
m_g^2 = \frac{m_s^2v_{\phi^*}G_{\it eff}}{\tilde{R}},\label{gmass}
\end{equation}
we find that the gravitons acquire a mass of $m_g$. In the scenario, the massive gravitational background scalar field $\phi^*$ modifies the Newtonian potential of DNS and contributes to a Yukawa-like one $\sim \frac{e^{-m_sr}}{r}$. The Yukawa-corrected potential has influence on the propagations of tensor gravitational waves, via the entrance of massive scalar component into the e.o.m. of gravitational waves. As a consequence, the massive gravitational background scalar field and two external scalar fields, manifested as scalar charges in the effective gravitational constant, have been eaten by the massless gravitons, remaining healthy massive gravitons with five d.o.f.. The gravitational background scalar fluctuation field $\sigma$ is the only massless field, which plays the role of Higgs-like field.

\section{Application to detected DNS binaries}

According to the expressions of (\ref{VEV}) and (\ref{smass}), we rewrite the mass of gravitons, Eq.(\ref{gmass}), as
\begin{equation}
m_g^2 = (-\alpha)^{3/2}\mu^3(2\lambda)^{-1/2}G_{\it eff}\tilde{R}^{-1}.\label{rgmass}
\end{equation}
It is found that the mass of graviton depends on three quantities\footnote{Noting that the self-coupling constant $\lambda$ of gravitational scalar field is of the order of unity.}, i.e. the separation of DNS system represented by $\tilde{R}$, the coupling strength between gravitational scalar field and the NSs that is characterized by the dimensionless coupling constant $\alpha$, and the scalar charges that is related to both the Planck scale constant $\mu$ and the effective gravitational constant $G_{\it eff}$. Consequently, the mass of gravitons rests with the intrinsic properties of DNS, i.e. the separation of the binaries and the compactness of two NSs, which is not a certain value and mildly variable.

It was proven that nonperturbative strong-gravitational-field effects developed in NSs for a dimensionless coupling constant $\alpha \le -4$, which induced order-of-unity deviations from general relativity \cite{1993PhRvL..70.2220D}. The general properties of binary systems consisting of scalarized NSs can be described by $\alpha \ge -4.5$, according to binary-pulsar measurements \cite{1998PhRvD..58d2001D, 2012MNRAS.423.3328F, 2013Sci...340..448A}. For $\alpha \le -5$, NSs in binary pulsar, with mass of 1.4 $M_{\odot}$, would develop strong scalar charges even in absence of external scalar solicitation, and a more negative value of $\alpha$ corresponds to a less compact NS \cite{1996PhRvD..54.1474D}. Most of the measured more massive NS in detected DNS systems have masses of $\sim 1.3-1.44 M_{\odot}$ \cite{2011A&A...527A..83Z}. Consequently, the coupling constant locates in a range of $\alpha = -5 \sim -6$ within a quadratic coupling model described in Eq. (\ref{tpotential}) \cite{1996PhRvD..54.1474D}. The scalar charges mildly vary with the compactness of NSs \cite{2014PhRvD..89d4024P} and will be $\sim 1$ only in the last stages of the evolution of NS binaries or close transient encounters. For NSs in the 8 detected DNS systems, the scalar charges are around 0.2 within solar-system bound \cite{1975ApJ...196L..59E} in the Fierz-Jordan-Brans-Dicke (FJBD) theory, by considering its dependence on the "sensitivities" $s \sim 0.2$ \cite{1996PhRvD..53.5541D, 2013PhRvD..87h4070M}. Accordingly, the gravitons radiated in a DNS binary with a semi-separation of $10^9$ m have masses of the order of $\sim 10^{-23} ev/c^2$, which mildly vary with the compactness of NSs and the separation between them (table \ref{tab:gmass}). The gravitons radiate from more closer DNS binaries possess a higher mass, which corresponds with current simulations that higher-frequency gravitational waves come from the closer binaries.

\begin{table*}
\begin{center}
\caption{Effective masses of gravitons radiated from 8 detected DNS binaries in our galaxy.}
\begin{tabular}{llllll}
\hline\hline
Source \ & \ $m_1$ \ & \ $m_2$ \ & \ $P_{orb}$ \ & \ Ecc \ & \ $m_g$ \\
  \ & \ ($M_{\odot}$) \ & \ ($M_{\odot}$) \ & \ (day) \ & \  \ & \ ($10^{-23} ev/c^2$) \\
\hline\hline
PSR J1811-1736 \ & \ $1.5^{+0.12}_{-0.4}$ \ & \ $1.06^{+0.45}_{-0.1}$ \ & \ 18.8 \ & \ 0.828 \ & \ 0.106 \\
\hline
PSR J1829+2456 \ & \ $1.35^{+0.46}_{-0.15}$ \ & \ $1.15^{+0.1}_{-0.25}$ \ & \ 1.176 \ & \ 0.139 \ & \ 0.711 \\
\hline
PSR J1913+16 \ & \ $1.44 \pm 0.0006$ \ & \ $1.39 \pm 0.0006$ \ & \ 0.323 \ & \ 0.617 \ & \ 0.972 \\
\hline
PSR B1534+12 \ & \ $1.35 \pm 0.0020$ \ & \ $1.33 \pm 0.0020$ \ & \ 0.421 \ & \ 0.274 \ & \ 1.292 \\
\hline
PSR B2127+11C \ & \ $1.36 \pm 0.080$ \ & \ $1.35 \pm 0.080$ \ & \ 0.335 \ & \ 0.681 \ & \ 1.408 \\
\hline
PSR J1756-2251 \ & \ $1.40^{+0.04}_{-0.06}$ \ & \ $1.18^{+0.06}_{-0.04}$ \ & \ 0.320 \ & \ 0.181 \ & \ 1.574 \\
\hline
PSR J1906+0746 \ & \ 1.37 \ & \ 1.25 \ & \ 0.166 \ & \ 0.085 \ & \ 2.427 \\
\hline
PSR J0737-3039 \ & \ $1.34 \pm 0.010$ \ & \ $1.25 \pm 0.010$ \ & \ 0.102 \ & \ 0.088 \ & \ 3.356 \\
\hline\hline
\end{tabular}
\label{tab:gmass}
\end{center}
{\bf Notes.}~~$m_1$ and $m_2$ are the masses of spontaneous scalarized NS and induced scalarized NS in units of solar mass, respectively. $P_{orb}$ denotes the orbital period in units of days. "Ecc" is the eccentricity for each binary system.
\end{table*}

\acknowledgments
This work is supported by the Fundamental Research Funds
for the Central Universities (Grant no. 161gpy49) at Sun Yat-
Sen University.

\end{document}